# Loss tangent measurements of dielectric substrates from 15 K to 300 K with two resonators: investigation into accuracy issues

Janina E. Mazierska[1), 2)], Mohan V. Jacob[1)], Dimitri O. Ledenyov[1)] and Jerzy Krupka[3)]

1) Electrical and Computer Engineering Department, School of Engineering, James Cook University, Townsville, QLD4811, Australia.
2) Massey University, Institute of Information Sciences and Technologies, Palmerston North, New Zealand.
3) Instytut Mikroelektroniki i Optoelektroniki Politechniki Warszawskiej, Koszykowa 75, 00-662, Warszawa, Poland.

The loss tangent of medium, low and very low loss dielectric substrates (including the *Rogers RT Duroid 5880* and *6010.2*, *LaAlO₃*, *(La, Sr)(Al, Ta)O₃*, *MgO* and *Quartz*) was measured at varying temperatures with two $TE_{01\delta}$ dielectric resonators to ensure verification of the tests. The accuracy of the measurements has been researched and discussed for split post dielectric resonator (*SPDR*) in a copper enclosure and a single post dielectric resonator (*SuPDR*) in a superconducting enclosure in the temperature range from *15K* to *300 K*.

## Introduction

Complex permittivity of dielectric substrates can be measured with the highest precision (currently), using the split post dielectric [1-4] and split cavity resonators [2]. Less precise methods include the microstrip and stripline resonator techniques. In practice, the data measured for the samples of cylindrical shapes is often assumed as well as the room temperature data, even if the dielectrics operate at the cryogenic temperatures. However, there are some differences in values of the complex permittivity between the bulk dielectrics and the planar dielectrics, and often the $\varepsilon_r$ is temperature dependent. Therefore, the above approximations may lead to erroneous prediction of microwave performance of devices, manufactured with a given planar material.

Designing a test resonator for dielectrics of differing complex permittivities is a challenging task, and it is difficult to obtain the same measurement accuracy for all materials. Also, while the random accuracy can be easily computed, the assessment of absolute accuracy is more complex as there are no perfect values available for comparison, especially at cryogenic temperatures.

To be able to verify our measurements of the loss tangent, we have constructed a set of two resonators: split post dielectric resonator (*SPDR*) in a *Copper* enclosure and a single post dielectric resonator (*SuPDR*) in a superconducting enclosure. We measured several dielectric substrates, namely the *Rogers RT Duroid 5880* and *6010.2*, *LaAlO₃*, *(La, Sr)(Al, Ta)O₃*, *MgO* and *Quartz*) with the both resonators and researched the measurement uncertainties in the loss tangent. Results of our investigations are present below and indicate that, in some cases, the two measurement fixtures should be used to ensure the reasonable accuracy of measurements.

## Measurement system

The measurement system consisted of a close cycle cryocooler *APD-HC4*, a vacuum dewar, a Vector Network Analyser *HP 8722C*, Temperature Controller *LTC-10*, a computer (*PC*), and the two dielectric resonators: split post dielectric resonator (*SPDR*) and single post dielectric resonator (*SuPDR*). The *SPDR* has been designed for measurements in a very wide range of temperatures from *20 K* to *300 K* and the highest resolution of $2 \times 10^{-5}$ at the temperature of *20 K*. The *SuPDR's* resolution reaches $2 \times 10^{-6}$ at the temperature of *15K*, and hence, it provides more accurate measurements, but only up to $T = 82\ K$.

a)
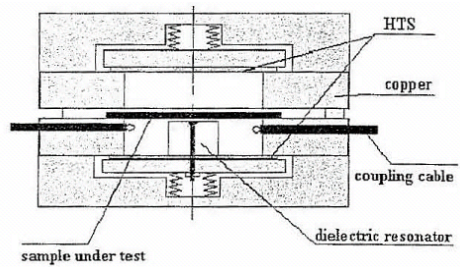

b)
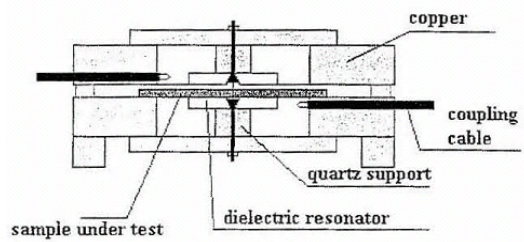

*Fig. 1. Schemes of (a) single post dielectric resonator (SuPDR) in a superconducting enclosure and (b) split post dielectric resonator (SPDR).*

The schemes of (*a*) single post dielectric resonator (*SuPDR*) in a superconducting enclosure and (*b*) split post dielectric resonator (*SPDR*) in a copper enclosure are shown in Fig. 1.

In our previous research paper, we presented the measurements of the real part of complex conductivity of low loss dielectrics. In this research paper, we discuss the measurement issues of the *tang δ*, computed with the use of the *Rayleigh-Ritz* model based analysis as in [4] from the *Q*-factors of empty resonators and with the substrates under test. The *Q*-factors were computed from the multi-frequencies measurements of the *S*-parameters around the resonance. The transmission mode *Q*-factor technique [4] was used to remove the noise, un-calibrated cables, adaptors and crosstalk from the measured *S*-parameters to ensure the high accuracy of measurements.

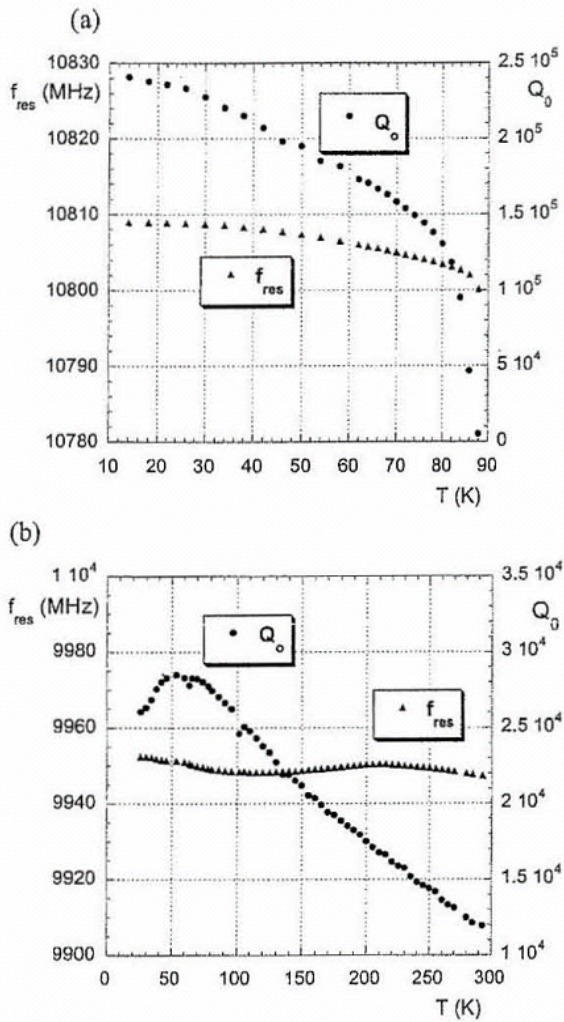

*Fig. 2. Unloaded $Q_0$-factor and resonance frequency $f_{res}$ of (a) superconducting post dielectric resonator and (b) split post dielectric resonator.*

### Experimental measurements results

Several medium, low and very low loss substrates, including the *Rogers RT Duroid 5880* and *6010.2*, *LaAlO$_3$*, *(La, Sr)(Al, Ta)O$_3$*, *MgO* and *Quartz* have been tested at varying temperatures from *15 K* to *300 K*. Thickness of the substrates was *0.25 mm*, *0.27 mm*, *0.500 mm*, *0.513 mm*, *0.508 mm*, and *0.400 mm* respectively. Measured values of temperature dependence of *tan δ* together with the measurements uncertainties $\Delta_r \, tan \delta$ are shown in Figs. 3-5. The measurements uncertainty $\Delta_r \, tan \delta$ has been calculated, using the principle of error in the difference between the two measured variables

$$\Delta_r \tan \delta = \left| \frac{\Delta \rho_e}{\rho_e} \right| \% + \frac{\sqrt{\left(\frac{\Delta Q_0}{Q_0}\right)^2 + \left(\frac{\Delta Q_D}{Q_D}\right)^2}}{\frac{1}{Q_D} - \frac{1}{Q_0}} \cdot 100\%$$

where $Q_0$ and $Q_D$ are the measured unloaded *Q*-factors of each resonator without and with a sample under the test, and $p_e$ is the energy filling factor of a given fixture. The *Q*-factor uncertainty of *5 %* (due to the cavity re-assembling) was used in the computations.

Measured losses of the **Rogers RT Duroid 5880** (Fig. 3a) and **6010.2** (Fig. 3b) substrates (shown for the first time in a wide range of temperatures) are between *6.8x10$^{-4}$* and *1.2x10$^{-3}$*; and *4.8x10$^{-4}$* and *9.7x10$^{-4}$* respectively.

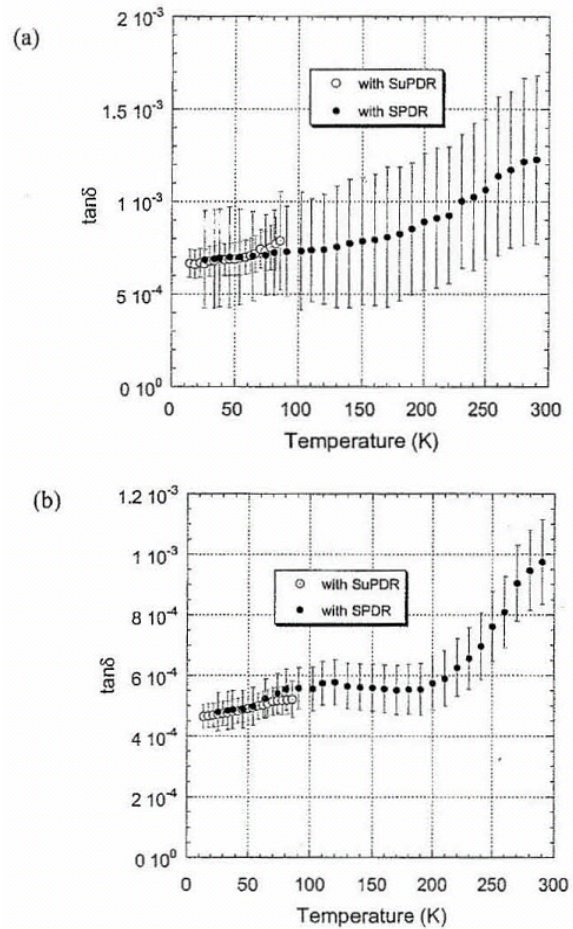

*Fig. 3. Loss tangent of Rogers RT Duroid (a) 5880 substrate at frequency of 9.9 GHz and (b) 6010.2 substrate at frequency of 9.6 GHz.*

Results obtained using the both resonators (illustrated in Figs. 3a and 3b) are very close, and the discrepancies are well within the measured uncertainties.

Tested **LSAT** substrates exhibited a maximum of the loss of *5.63x10⁻⁴* at the temperature of *150 K* (Fig. 4) as measured with the *SPDR*. The results between the temperatures of *50 K* and *82 K* were confirmed with the *SuPDR* with the discrepancies below *13 %*.

Values on *tanδ* below the temperature of *50 K*, measured with the *SPDR*, were bigger than those with the *SuPDR* by up to *78 %* at *20 K* due to the finite resolution of the *SPDR*. Measurements with the two resonators have allowed for obtaining the correct temperature dependence in a wide range of temperatures, while with the *SPDR* only would have resulted in a false minimum at *50 K*.

For **LAO** substrates (Fig. 5), the strange phenomenon was observed, namely the results with the high resolution fixture (*SuPDR*) were much higher that with the lower resolution resonator. This was initially very surprising, however the results can be explained, if the electric non-uniformity of the *LAO* substrates and the differences in the dielectric constants of the *SPDR's* and *SuPDR's* rods are taken into the consideration. Therefore, the only use of the *SuPDR* for the *LAO* samples testing would have resulted in almost *50 %* too high values of the perpendicular component of loss tangent.

Measured values of *tanδ* of **MgO** (shown in Fig. 6) and **Quartz** (Fig. 7) show the big differences with the *SPDR*'s measurements resulting in much higher values than the *SuPDR*. This was expected as the losses of the both types of samples under the test are below the resolution of the *SPDR*. However, the bottom error band of the *SPDR*'s results is within the error bands of the *SuPDR*. Hence, the *SPDR* results can be used to determine the upper bound values of loss tangent only.

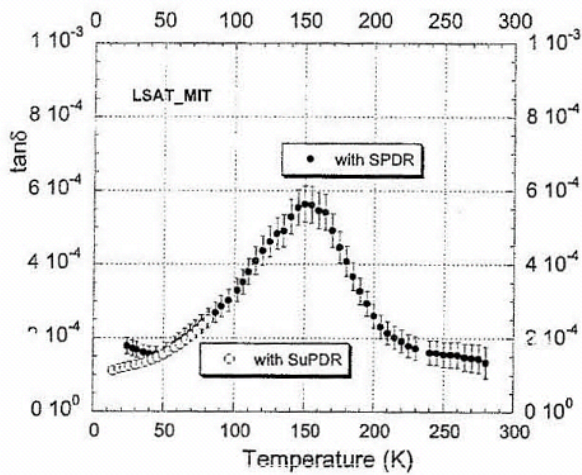

*Fig. 4. Loss tangent of LSAT substrate at frequency of 8.68 GHz.*

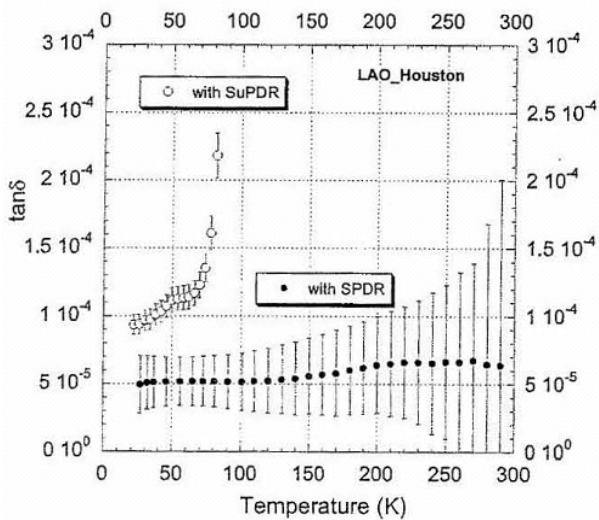

*Fig. 5. Loss tangent of LAO UoH substrate at frequency of 8.58 GHz.*

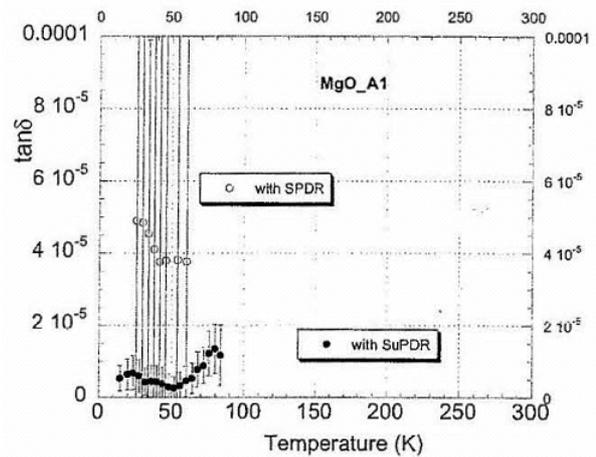

*Fig. 6. Loss tangent of MgO substrate at frequency of 9.42 GHz.*

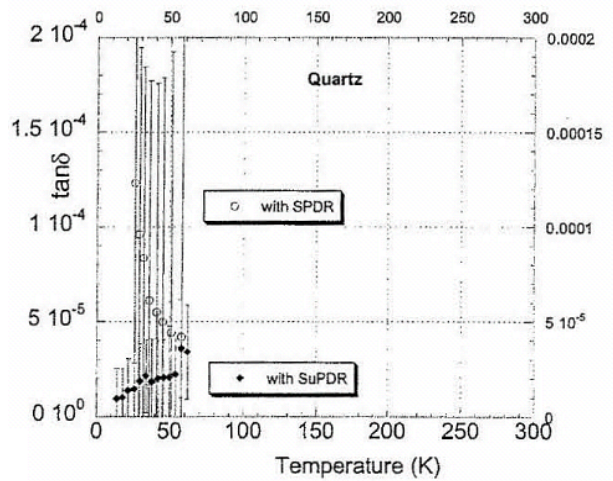

*Fig. 7. Loss tangent of Quartz substrate at frequency of 9.8 GHz.*

## Conclusions

Microwave characterization of planar low loss dielectrics of varying properties and varying temperatures, performed with a single resonator of even high resolution, may result in the low accuracy of measured perpendicular loss tangent for some materials. For medium loss substrates like the *RT Duroid 5880*, and *6010.5* (with the *tanδ* bigger than $4 \times 10^{-4}$), using one resonator (*Split Post Dielectric Resonator*) only was sufficient to provide the accurate results of losses for the temperatures from the cryogenic temperatures up to the temperature of *300 K*.

However, for the substrates with the *tanδ* smaller $5 \times 10^{-5}$ (*LAlO₃*, *LSAT*, *MgO* and *Quartz*), adding the second *SuPDR* resonator, working in a narrow range of temperatures, but of very high resolution of $2 \times 10^{-6}$, turned out to be necessary to enable the verification (and the elimination of some results) of wide temperature measurements fixture. The two resonators system has proved to allow for more accurate microwave measurements of the planar dielectric of medium and low loss in a wide temperatures range from *20 K* to *300 K* than a single measurement fixture.

Measurements of very low loss substrates like the *MgO* (*tanδ* $< 2 \times 10^{-5}$), using the *SPDR*, resulted in the very high values of the *tanδ* due to its low resolution. Therefore, in the cases of *MgO* and *Quartz*, the *SuPDR* should be used. However, we can assess the upper bound loss, by subtracting the measurement uncertainty from the measured values, when using the *SPDR*.

## Acknowledgement

Acknowledgements: this research work was partly sponsored by Massey University, *ARC* Linkage Grant *LX0561280*, and *ARC* Discovery Grant *DP0449996*.

This research paper was published in the *Proceedings of Asia Pacific Microwave Conference 2005* in [13].

*E-mail: janina.mazierska@jcu.edu.au

———————